\documentclass[a4paper,11pt]{article}
\usepackage{amsmath,amssymb,amsfonts,latexsym,comment}
\usepackage{mathrsfs}
\usepackage{moreverb}
\usepackage{graphicx}
\usepackage[english,activeacute]{babel}
\usepackage{multicol}
\usepackage{hyperref}
\usepackage{tikz} 
\usepackage{pgfplots}  
\usepackage{algorithmic}
\usepackage{algorithm}
\usepackage{mathtools}

\DeclarePairedDelimiter\floor{\lfloor}{\rfloor}

\newcommand{\triplesum}{\sum_{{\bf \kk} \in \KK }}
\newcommand{\kk}{\bf \gamma}
\newcommand{\KK}{\Gamma}

\newcommand{\N}{\mathbb{N}}

\newcommand{\M}{\mathcal{M}}

\newcommand{\Per}{{\mathcal P}}

\newcommand{\List}{{\mathcal L}}
\newcommand{\C}{{\mathcal C}}
\newcommand{\Steg}{{\mathcal S}}
\newcommand{\fracd}[2]{\displaystyle{\frac{{\displaystyle{#1}}}{{\displaystyle{#2}}}}}
\newcommand{\fracsum}[2]{#1 \; / \; #2 }

\begin{document}

\title{A steganographic approach based on the chaotic fractional map and in the DCT domain}
\author{A. Soria-Lorente, E. P\'erez-Michel, E. Avila-Domenech\\
Tecnology Department\unskip, 
University of Granma\unskip, Bayamo\unskip, 85100\unskip, Granma\unskip, Cuba\\
asorial@udg.co.cu, asorial1983@gmail.com}
\date{\emph{(\today)}}
\maketitle

\begin{abstract}
A steganographic method based on the chaotic fractional map and in the DCT domain is proposed. This method embeds a secret message in some high frequency coefficients of the image using a 128-bit private key and a chaotic fractional map which generate a permutation indicating the positions where the secret bits will be embedded. An experimental work on the validation of the proposed method is also presented, showing performance in imperceptibility, quality, similarity and security analysis of the steganographic system. The proposed algorithm improved the level of imperceptibility and Cachin's security of stego-system analyzed through the values of Peak Signal-to-Noise Ratio (PSNR) and the Relative Entropy (RE).

\medskip


\medskip


\medskip

\end{abstract}

\section{Introduction}
\label{intro}

With the development of the technologies of information technology and communications and the rapid growth of network bandwidth, the Internet has turned out to be a much-used channel for transmitting much information in diverse audio, video, image, and text digital form. This information can be public or private, in case of private information; it may increase the risk to reveal this information by unauthorized external factors. Thus, there is a real necessity to safeguard sensitive information against access and illegal manipulation. Currently, the Cryptography and Steganography offer the level of security required for private communication against human interception. Cryptography provides security to the content of the message whereas steganography conceals the existence of the message itself\cite{SubhedaraSH}. Modern Steganography techniques offer such security levels. Steganography is the art and science of secret communication to conceals sensitive information from an intermediary \cite{ArunaMH,majumder2013}. In addition, in the steganography, the secret message is embedded inside of several cover media, such as text \cite{din2019evaluation,gutub2018multi,naqvi2018multilayer,vaishakh2019semantic}, image \cite{hussain2018image,liao2018medical,nipanikar2018sparse,zou2019image}, video \cite{alzain2018application,chen2018novel,manisha2018two} and audio \cite{mishra2018audio,zhang2019audio} files. There are two main criteria by which the performance of steganographic algorithms can be measured, embedding capacity and detectability. Therefore, new steganographic algorithms are expected to increase the image capacity and the encryption strength of the message. Image capacity can be increased by adaptive strategies that decide where it is better to embed the secret message. In steganography, the embedding capacity is defined as the largest sequence of bits that can be embedded in a given cover image \cite{SoriaSCN}.

\subsection{Literature survey}
In recent years, several digital steganography techniques have been proposed. All of them share the idea of inserting secret information into a cover media to generate a stego output. There are fundamentally two types of steganographic techniques: techniques in the spatial domain and in the frequency domain \cite{SinghPS}. In the frequency domain, are used some transforms, such as the discrete cosine transform (DCT).

In \cite{el_rahman2016comparative}, Sahar and Rahman developed a steganographic tool based on the Discrete Cosine Transform to embed confidential information about a nuclear reactor. In particular, this tool uses the method of sequential embedding in the mean frequency of the image. On the other hand, in \cite{attaby2017data} a new steganographic technique in JPEG images in the frequency domain is shown, which provides high embedding performance while introducing minimal changes to the cover carrier image. This algorithm uses module three of the difference between two DCT coefficients to embed two bits of the compressed form of the secret message.

Recently, in \cite{singh2018}, the authors described a steganographic method based on DCT and the entropy threshold technique. This steganographic algorithm uses a random function to select in which block of the image the bits extracted from the binary sequence of a secret message will be inserted. Moreover, in another paper, Chowdhuri and others present a steganographic scheme based on a matrix weighted for a color image with a high degree of through the to a balance between and imperceptibility. In the paper, AC components are collected from matrices ($8\times 8$) of quantified DCT coefficients of the YCbCr channel. Then, a set of matrices ($3\times 3$) are formed to hide secret data. A shared 128-bit secret key controls the gathering of AC components. Finally, the authors demonstrate experimentally that the proposed scheme provides good payload and high visual quality compared to existing state-of-the-art methods \cite{chowdhuri2018secured}.

Chaotic maps are being used in recent years to increase the security of digital steganographic systems. In secure communications, chaotic maps have some characteristics such as; extreme sensitivity to the initial conditions, the expansion of the orbits throughout the space, random behavior, control parameters, and ergodicity. These specific properties make chaotic maps excellent candidates for steganography and encryption. Hence, some applications of these maps can be found in the literature, for example:

In \cite{habib2015enhancement}, a secure steganographic method based on the frequency domain is proposed. It allows you to hide a secret image in another image randomly using Chaos. The chaotic generator Peace Wise Linear Chaotic Map with disturbance was selected, it has good chaotic properties and it is easy to implement. To obtain the pseudo-random pixel sequence in which the secret image will be embedded in its DCT coefficients the chaotic generator was used. Experimental results show that the proposed algorithm achieves high quality and security. Another case it is presented in \cite{SaidiHRB}, the authors describe a new steganographic scheme in the frequency domain based on a chaotic map. In the proposed scheme, they apply DCT on the cover image and collect the AC coefficients in zigzag order. The process of embedding and extracting the secret message depends on a Piecewise Linear Chaotic Map \cite{LiChM}, where its initial condition and control parameters are adopted as secret keys of the designed scheme. 

\subsection{Motivation and objective}

Guaranteeing the security of long-distance communication is a critical problem. This is particularly important in the case of storage and transmission of confidential data in a public network such as the Internet. The security of such data communication that is mandatory and vital to many current applications has been a primary concern, and an ongoing issue since the Internet is open in nature and public by plan \cite{yahya2018steganography}.

It is remarkable that it is impossible to obtain the maximum embedding capacity with an acceptable level of imperceptibility at the same time. So, conciliation must be made between imperceptibility and embedding capacity. For different applications, the acceptable equilibrium between these two constraints is diverse, depending on the nature of the requirements of the application.

Thus, in this research, the objective is to reach maximum possible level of imperceptibility by keeping highest embedding capacity and Cachin's security of the stego-system.

In this paper, a secure and imperceptible DCT steganography method is proposed. It allows embeds a secrete message in the first eight AC coefficients of high frequency of a cover image, taking into account a 128-bit private key and Chaos Theory. The chaotic generator Fractional Chaotic Map was selected. It was used to obtain the pseudo-random position of the previous collected AC coefficients in which the secrete message will be embedded following LSB substitution. Then, the stego-image is reconstructed. Moreover, its shown an experimental comparison of the proposed method with respect to existing state-of-the-art methods (Sahar and Rahman Method \cite{el_rahman2016comparative}, Chowdhuri et al. Method \cite{chowdhuri2018secured}., Habib et al. Method \cite{habib2015enhancement}, and Saidi et al. Method \cite{SaidiHRB}).  Finally, the proposed scheme provides high imperceptibility level, strong security and good embedding capacity  compared to existing state-of-the-art methods mentioned above. 

In addition, the proposed algorithm is implemented in Python 3.7. For the experimental analysis several color images with size (512$\times$512) were collected from two different datasets: image dataset of 1500 RGB-BMP images, transformed from Caltech birds' dataset in JPEGC format \cite{AlJarrahM} and image dataset of 1500 RGB-BMP images, transformed from NRC dataset in TIFF format \cite{AlJarrahM}. For the experimental results several different randomly generated keys were used. The proposed algorithm has been tested by inserting a message of 98304 bits in length.

The structure of the document is as follows: In Section 2, shows the mathematical background. Section 3 defines the proposed scheme. Results and discussion are described in Section 4. Finally, the conclusions arrived.

\section{Mathematical background}
\subsection{Discrete Cosine Transform}
Let $\C$ be the cover image (gray scale or RGB images) and let $\mathcal{K}$ be the set of all the non-overlapping $8\times8$ bytes blocks of $\C$, such that
\begin{equation}
\C = \bigcup_{k\in\mathcal{K}} B^{k}.\label{Cover_Bloks}
\end{equation}
Moreover, let $\mathcal{B}^{k}$ be the two dimensional discrete cosine transform. The relationship between $\mathcal{B}^{k}\equiv\mbox{DCT}$ and its inverse $B^{k}\equiv\mbox{IDCT}$ (Inverse Discrete Cosine Transform) is given by
\begin{equation}
\mathcal{B}^k_{u,v}= 4^{-1}\sigma(u)\sigma(v) \sum_{0\leq	i,j\leq 7}B^k_{i,j}\cos\left(\fracd{\pi u(2i+1)}{16} \right)\cos\left(\fracd{\pi v(2j+1)}{16} \right),\label{DDCT}
\end{equation}
to each selected block. Here $0\leq u,v\leq 7$ and $\sigma(x) = \sqrt{2^{-1}}$ for $x=0$ and $\sigma(x) = 1$ otherwise.

In addition, the IDCT is given by
\begin{equation}
B^k_{i,j}=4^{-1}\sum_{0\leq
	u,v\leq 7}\sigma(u)\sigma(v)\mathcal{B}^k_{u,v}\cos\left(\fracd{\pi u(2i+1)}{16} \right)\cos\left(\fracd{\pi v(2j+1)}{16} \right), \label{IDCT}
\end{equation}

\subsection{Quantification procedure and zigzag scan order}
From \eqref{DDCT} each integer block $B^k$ is transformed into a real block $\mathcal{B}^k$, which is scaled according to the quantification matrix $Q^{\mu}$ by a compression quality factor $\mu$
\begin{equation}
Q^{\mu}=\chi(\mu)\begin{pmatrix}
16 & 11 & 10 & 16 & 24 & 40 & 51 & 61 \\ 
12 & 12 & 14 & 19 & 26 & 58 & 60 & 55 \\ 
14 & 13 & 16 & 24 & 40 & 57 & 69 & 56 \\ 
14 & 17 & 22 & 29 & 51 & 87 & 80 & 62 \\ 
18 & 22 & 37 & 56 & 68 & 109 & 103 & 77 \\ 
24 & 35 & 55 & 64 & 81 & 104 & 113 & 92 \\ 
49 & 64 & 78 & 87 & 103 & 121 & 120 & 101 \\ 
72 & 92 & 95 & 98 & 112 & 100 & 103 & 99  %
\end{pmatrix} \label{QTable}
\end{equation}
where
$\displaystyle\chi(\mu) = \frac{100-\mu}{50}$, with $50<\mu<100$.

\begin{figure}[H]
	\centering\includegraphics[scale=.35]{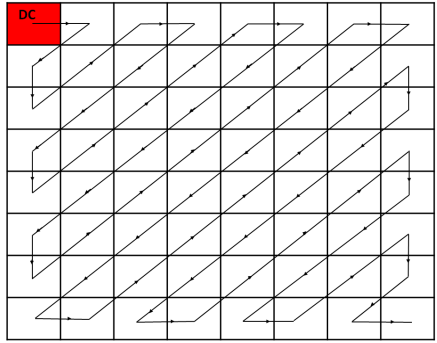}
	\caption[]{Zigzag order scan}
	\label{ZZagF}	
\end{figure}

The process of quantization is accomplished by dividing each element $\mathcal{B}^k$ by the corresponding element in the quantification matrix $Q^{\mu}$, and then rounding to the nearest integer, i.e., the coefficients of the quantized blocks $\varTheta^{k}_{u,v}$ are computed by
\begin{equation}
\varTheta^{k}_{u,v}=\text{round}\left(\fracd{\mathcal{B}^{k}_{u,v}}{Q_{u,v}^{\mu}}\right) ,\quad 0\leq u,v\leq 7.\label{quantification}
\end{equation}
Next, it applies the zigzag scan to the matrix of the quantized coefficients $\left(\varTheta^{k}_{u,v}\right)_{0\leq u,v\leq 7}$, see Figure \ref{ZZagF}, with the purpose of aligning frequency coefficients in ascending order, starting from frequency zero (DC coefficient) to high frequency components (AC coefficients) see \cite{SoriaSCN,ZhuH2}.
On the other hand, the unquantization process for $\overline{\varTheta}^{k}_{u,v}$ is carried out by
\begin{equation}
\overline{\mathcal{B}}^{k}_{u,v}=\text{round}\left(\overline{\varTheta}^{k}_{u,v}Q_{u,v}^{\mu}\right),\quad 0\leq u,v\leq 7.\label{Dq}
\end{equation}

\subsection{Chaotic permutation. Fractional chaotic map}

Chaotic maps have been used to increase the digital security of steganography. Edward Lorenz discovered the first chaotic system in 1963 \cite{lorenz1995essence}. Since then, different areas of research in engineering, physics and mathematics have been established. The most important specification of chaos is its sensitivity to initial conditions. If the primary is different but very similar conditions are chosen, the output of the system will be very different and a pseudo-random string will be produced. In steganography, chaotic maps are used to find the insertion positions in the cover image and the bit positions using the least significant bit method. The chaotic signals sound like noise but are completely safe. This means that if the primary values and the map function are given, the value of the signal can be reproduced again \cite{GagnaniAndVarjani2015}.

Recently, several authors \cite{valandar2018integer,walia2018robust,yadav2018chaotic} have proposed embedding schemes based on chaotic maps, which are nonlinear system, characterized by a pseudo random behavior and an high sensitivity to initial conditions and control, unpredictability, ergodicity, etc \cite{martinez2016steganographic}. In this research, it is used the fractional chaotic map proposed in \cite{bai2018novel}, which is defined as follows
\begin{equation}
x(n+1)=x(0)+\frac{1}{\Gamma(\nu)}\sum_{0\leq j\leq n}\frac{\Gamma(n-j+\nu)}{\Gamma(n-j+1)}g(j,x(j)),\quad 0 < \nu\leq 1,
\end{equation}
where $\Gamma(x)$ is the gamma function \cite{andrews1992special} and $g(n,x(n))$ is defined in \cite{bai2018novel}. Thus, the chaotic permutation of $\varrho=\{\varrho_1,\ldots,\varrho_n\}$ is determined by
\begin{equation}
(\varrho_j)_{j\in\Per},\quad\mbox{where}\quad \Per=\left\lbrace \floor*{x(i)10^{14}\mbox{ mod. }n},\quad 1\leq i\leq n\right\rbrace . \label{fcmg}
\end{equation}

\subsection{Statistical measures to compute the image quality and security}

\subsubsection{Peak Signal to Noise Ratio (PSNR)}
As the cover image is altered to embed the secret data, there will be changes in the cover image pixel values. Thus, that the changes need to be analyzed since it directly affect the imperceptibility of the output stego image. Peak Signal to Noise Ratio (PSNR) is one of the popular and top notch metric used to measure the quality of the stego image by analyzing the mean squared error value between the cover and the stego image \cite{kadhim2018comprehensive}. Indeed, this measure is used to evaluate the invisibility of a secret message \cite{Atta2018} as well as the imperceptibility \cite{AAwad} and the visual quality \cite{sahu2019novel,ValandAB} of the stego image compared to the cover image, wstatistical measures compute the image qualityith decibel (db) as measurement unit. Higher PSNR indicates that the reconstruction of the image is of higher quality, \cite{DattaBK}. The PSNR is given by \cite{SoriaSCN}
\begin{equation*}
\text{PSNR}=10\log _{10}\left( \fracd{\Xi^{2}}{\text{MSE}}\right) ,
\end{equation*}
where
\begin{equation*}
\text{MSE}=(mn\rho)^{-1}\triplesum
\left\Vert
\mathcal{C}\left( \kk \right) -\mathcal{S}\left( \kk \right) \right\Vert ^{2},
\end{equation*}
and $\mathcal{C}$ and $\mathcal{S}$ are the cover image and the stego image respectively, of size $m\times n\times \rho$, with $\mathcal{C}, \mathcal{S} \in \{0, 1, \dots, \Xi \}$, and $\Xi=\text{max}(\text{max}(\mathcal{C}),\text{max}(\mathcal{S}))$.

The index set ${\kk} = (\ell_1, \ell_2, \ell_3)$ sums over the set
\begin{align*}
{\KK} = \{1, \dots, m \} \times \{ 1, \dots, n \} \times \{1, \dots, \rho \},
\end{align*}
where $\rho=1$ for gray scale images and  $\rho=3$ for 24-bit color images.

\subsubsection{Universal Image Quality Index (UIQI)}
Usually the image quality based on the Human Visual System (HVS) is measured by the Universal Image Quality Index (UIQI), which was proposed by Wang and Bovik in \cite{WangBo}. This measure is universal in the sense that it does not take the viewing conditions or the individual observer into account \cite{Bayra}. Moreover, it does not use traditional error summation methods \cite{ZhengYQ}. The dynamic range of UIQI is between -1 and 1. For identical images its value will be 1.
\begin{equation*}
\text{UIQI}=\fracd{4\sigma_{\mathcal{C}\mathcal{S}}}{\sigma_{\mathcal{C}}^2+\sigma_{\mathcal{S}}^2}\,\fracd{\overline{\mathcal{C}}\,\overline{\mathcal{S}}}{\overline{\mathcal{C}}^2+\overline{\mathcal{S}}^2},
\end{equation*}
where
\begin{equation*}
\overline{\mathcal{C}}=(mn\rho)^{-1}\triplesum
\mathcal{C}\left( \kk \right),
\end{equation*}
\begin{equation*}
\overline{\mathcal{S}}=(mn\rho)^{-1}\triplesum
\mathcal{S}\left( \kk \right),
\end{equation*}
\begin{equation*}
\sigma_{\mathcal{C}}^2=(mn\rho-1)^{-1}\triplesum
\left( \mathcal{C}\left( \kk \right)-\overline{\mathcal{C}}\right)^2,
\end{equation*}
\begin{equation*}
\sigma_{\mathcal{S}}^2=(mn\rho-1)^{-1}\triplesum
\left( \mathcal{S}\left( \kk \right)-\overline{\mathcal{S}}\right)^2,
\end{equation*}
\begin{equation*}
\sigma_{\mathcal{C}\mathcal{S}}=(mn\rho-1)^{-1}\triplesum
\left[\left( \mathcal{C}\left( \kk \right)-\overline{\mathcal{C}}\right)\left( \mathcal{S}\left( \kk \right)-\overline{\mathcal{S}}\right) \right] ,
\end{equation*}

\subsubsection{Image fidelity (IF)}
Image fidelity is a measure that shows a consistent relationship with the quality perceived by the human visual perception. Moreover, it is a metric that measure the similarity between the cover image $\mathcal{C}$ and the stego image $\mathcal{S}$ after insertion of the message without any visible distortion or information loss \cite{SoriaSCN}. It is defined by~\cite{Khamrui,SenguptaM,SoriaSCN}
\begin{equation*}
\text{IF}=1-\fracsum{\displaystyle 
	\triplesum
	\left( \mathcal{C} \left(  \kk \right) - \mathcal{S} \left( \kk  \right)\right) ^2}{\displaystyle
	\triplesum
	\mathcal{C} \left( \kk \right)^2} ,
\end{equation*}

\subsubsection{Relative entropy (RE)}
The security of a steganographic system is defined in terms of the relative entropy
\begin{equation*}
\text{RE}\left( P_{\mathcal{C}}||P_{\mathcal{S}}\right) =\sum P_{\mathcal{C}}\left\vert \log \fracd{P_{\mathcal{C}}}{%
	P_{\mathcal{S}}}\right\vert,
\end{equation*}%
where $P_{\mathcal{C}}$ and $P_{\mathcal{S}}$ represent the distribution of cover and stego image, respectively. This statistical measure was proposed by Cachin in~\cite{Cachin}. Moreover, a steganographic system is said to be
\begin{itemize}
	\item $\varepsilon$-secure if $\text{RE}\left( P_{\mathcal{C}}||P_{\mathcal{S}}\right)\leq \varepsilon$,
	\item perfectly secure if $\text{RE}\left( P_{\mathcal{C}}||P_{\mathcal{S}}\right) =0$.
\end{itemize}
Summing up, for the $\text{RE}\left( P_{\mathcal{C}}||P_{\mathcal{S}}\right)$, the closer the value is to 0, the higher the level of security.

\subsection{Blox plots}

\begin{figure}[H]
	\centering\includegraphics[scale=.49]{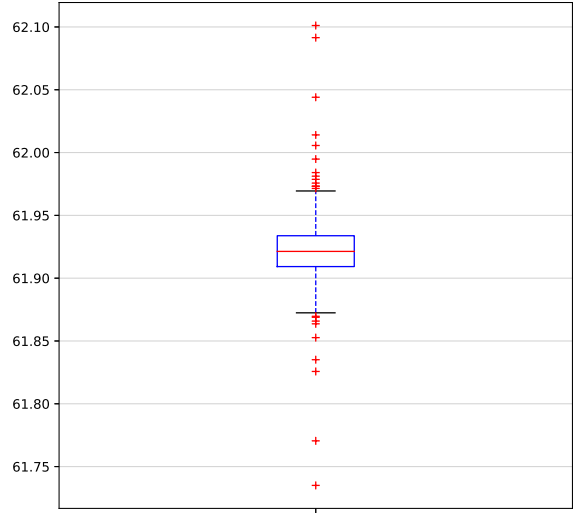}
	\caption[]{Blox plot} 
	\label{BoxP}	
\end{figure}

In the box plots drawn in Figure~\ref{BoxP}, the horizontal axis represents the different methods that are compared, and the vertical axis represents the PSNR values. The upper and lower limit of the rectangle are the upper and lower quartiles ($Q_1$ and $Q_3$) of test results separately, and the difference between the upper and lower quartile is the quartile difference IQR. The red line in the rectangle is the median. The two black horizontal lines at $Q_3 + 1.5\text{IQR}$ and $Q_1-1.5\text{IQR}$ are the cut-off points for abnormal values, known as the internal limit. The data outside the internal limits is outliers and is represented by the red ‘+’.

\subsection{Notations}
In this work, the following notations are taken into account:
\begin{itemize}
	\item $|\varpi|$ denote the number of elements of $\varpi$.\\
	
	\item $||$ denote the concatenation.\\
	
	\item $\Delta(\eta)$ denote the function that reorganizes the vector $\eta$ of length 64 to a matrix of order $8$, taking into account the zigzag scan order \ref{ZZagF}.\\
	
	\item R$(x,\beta)$ denote the function that replaces the Least Significant Bit (LSB) of $x\in\N$ by $\beta\in\{0,1\}$, see \cite{SoriaSCN}.\\
	
	\item R$^{-1}(x)$ denote the function that extracts LSB of $x\in\N$, see \cite{SoriaSCN}.\\
	
	\item $A\setminus B$ denote the set difference of $A$ and $B$.

\end{itemize}

\section{Proposed scheme}

In this Section its present a DCT steganographic algorithm, which use a 128-bit private key. In addition, it is assumed that the sender as well as the receiver hold the same system of 128-bit private keys. The secret message is inserted into the cover image $\mathcal{C}$ and is extracted from the stego image $\mathcal{S}$ by the embedding and extracting algorithm, respectively.

\subsection{Embedding algorithm}
The Algorithm \ref{alg:embedding} shows the step by step embedding process. Firstly, the cover image $\mathcal{C}$ is divided into non-overlapping $8\times8$ bytes blocks, see \eqref{Cover_Bloks}. Afterwards, DCT coefficients are obtained from each $8\times8$ blocks, see \eqref{DDCT}. The quantized DCT coefficients from each block using quantization matrix \eqref{QTable} are obtained. Then, the first eight AC coefficients, are collected from all the $8\times8$ DCT coefficient matrices and stored into a coefficient 1D-array $\omega$. The zigzag scan order is used to collect the coefficients, see Figure \ref{ZZagF}. In order to increase the security of the stego system, a 128-bit private key $\kappa$ is used. From this a 1024-bit sequence is generated using BLAKE2B \cite{aumasson2013blake2,sugier2017implementation}. Then, this sequence is expanded to a binary sequence $\overline{\kappa}$ with the same length of $\omega $. Next, $\omega =\omega^{(1)}||\omega^{(2)}||\cdots||\omega^{(i)}||\cdots $ and $\overline{\kappa}=\overline{\kappa}^{(1)}||\overline{\kappa}^{(2)}||\cdots||\overline{\kappa}^{(i)}||\cdots$ are partitioned into logical sequences of length 64. Then, for each $\omega^{(i)}$ the secret bits collected from secrete message $\mathcal{M}=\{m_\ell\in\{0,1\}\,:\, 1\leq \ell\leq |\mathcal{M}|\}$ are embedded in LSB of the value of the coefficient of a particular position $\rho_i$, obtaining a modified 1D-array $\overline{\omega}^{(i)}$. The $\rho_i$ positions are determined by $\overline{\kappa}^{(i)}$ and chaotic positions generator $(\varrho_i)_{1\leq i\leq 64}$ given by \eqref{fcmg}, see Algorithm \ref{Permutation}. Then, the coefficient 1D-array $\omega$ is reconstructed by collecting all the modified $\overline{\omega}^{(i)}$. After that, a
new $8\times8$ quantized DCT matrices are formed from inverse zigzag scan order of the 1D-array $\omega$. Applying inverse quantization \eqref{Dq} and IDCT \eqref{IDCT} to all $8\times8$ resulting matrices, stego pixel blocks are acquired and finally the stego image is generated.

\begin{algorithm}[H]
	\caption{ChaoticPositions}
	\label{Permutation}
	\begin{algorithmic}[1]	
		\STATE \text{{\bf Input:}} 	$\overline{\kappa}^{(i)}$, $x(0)$, $\nu$.
		\STATE \text{{\bf Output:}} $\rho=\{\rho_1,\ldots,\rho_{64}\}$	
		\STATE $\List=\{1,\ldots,64\}$	
		\STATE Given $x(0)$ and $\nu$, get chaotic permutation $\varrho=(j)_{j\in\Per}$ of $\List$ from equation \eqref{fcmg}\\		
		\mbox{\textbf{/*First collect the positions where the key has a bit set to one.*/}}
		\STATE $j=k=1$
		\FOR {\textbf{each} bit \textbf{in} $\overline{\kappa}^{(i)}$} 		
		\IF{current bit is equal to one}		
		\STATE $\rho_k=\varrho_j$		
		\STATE $k=k+1$		
		\ENDIF
		\STATE $j=j+1$
		\ENDFOR\\		
		\mbox{\textbf{/*Then, pick up the positions where the key does not have a bit set to one.*/}}
		\IF{ length$(\rho)$ less than 64}
		\STATE Given $x(0)$ and $\nu$, get chaotic permutation $\overline{\varrho}$ of $\varrho\setminus\rho$ from equation \eqref{fcmg}
		\STATE $\rho=\rho\cup\overline{\varrho}$
		\ENDIF
		\RETURN $\rho$
	\end{algorithmic} 
\end{algorithm}

\begin{algorithm}[H]
	\caption{Embedding Algorithm}
	\label{alg:embedding}
	\begin{algorithmic}[1]	
		\STATE \text{{\bf Input:}} Cover image $\mathcal{C}$, 128-bit private key $\kappa$, secret message $\M$, $x(0)$, $\nu$	
		\STATE \text{{\bf Output:}} Stego image $\Steg$\\
		\mbox{\textbf{/* Collecting AC coefficients */}}
		\STATE Divide $\C$ into $\mathcal{K}$ non-overlapping blocks of $8\times8$ bytes, see \eqref{Cover_Bloks}
		
		\STATE $\omega=\emptyset$
		
		\STATE $\ell=0$
		
		\FOR{\textbf{each} $B^k\in\C$}
		
		\STATE $\mathcal{B}^k\gets B^k$ : DCT($B^k$) according to~\eqref{DDCT}
		
		\STATE $\varTheta^{k} \gets \mathcal{B}^{k}$		
		: Quantify $\mathcal{B}^{k}$ according to~\eqref{quantification}
		
		\STATE $\nu^{k} \gets \varTheta^{k}$ : Apply the zigzag scan, see Figure \ref{ZZagF}
		
		\STATE $\omega\gets\omega\cup\{\nu_j^{k}\}_{2\leq j\leq 9} $
		
		\ENDFOR\\
		\mbox{\textbf{/* Key expansion using blake2b algorithm */}}
		\STATE From $\kappa$ generate 1024-bit sequence using BLAKE2B algorithm
		
		\STATE Expand the previous sequence to a binary sequence $\overline{\kappa}$ with the same length of $\omega$\\
		\mbox{\textbf{/* Embbeding secrete bits into collected AC */}}
		
		\STATE Partition $\omega \gets\omega^{(1)}||\omega^{(2)}||\cdots||\omega^{(i)}||\cdots $ and $\overline{\kappa}\gets\overline{\kappa}^{(1)}||\overline{\kappa}^{(2)}||\cdots||\overline{\kappa}^{(i)}||\cdots$ into a logical sequence of length 64
		
		\FOR{\textbf{each} $\omega^{(i)}\in\omega$}
		
		\STATE $\rho\gets\mbox{ChaoticPositions}(\overline{\kappa}^{(i)}, x(0), \nu)$
		
		\FOR{\textbf{each} $\psi\in\rho$}
		
		\STATE $\ell\gets\ell + 1$
		
		\IF{$\omega^{(i)}_{\psi}<0$}
		
		\STATE $\bar{\omega}^{(i)}_{\psi}\gets -$R$(\mbox{abs}(\omega^{(i)}_{\psi}), m_{\ell})$;
		
		\ELSE
		
		\STATE $\bar{\omega}^{(i)}_{\psi}\gets $R$(\omega^{(i)}_{\psi}, m_{\ell})$;
		
		\ENDIF
		
		\ENDFOR
		
		\STATE $\omega^{(i)}\gets\bar{\omega}^{(i)}$; Reconstruct 1D-array $\omega$ of coefficient from modified $\bar{\omega}$
		
		\ENDFOR\\
		\mbox{\textbf{/* Grouping AC coefficients and rebuilding the stego image */}}
		
		\STATE Partition $\omega \gets\varpi^{(1)}||\varpi^{(2)}||\cdots||\varpi^{(i)}||\cdots$ into a logical sequence of length 8
		
		\FOR{\textbf{each} $B^k\in\C$}
		
		\STATE $\{\nu_j^{k}\}_{2\leq j\leq 9}\gets\varpi^{(k)}$
		
		\STATE $\overline{\varTheta}^{k} \gets\Delta(\overline{\nu}^k)$
		
		\STATE $\overline{\mathcal{B}}^k\gets\overline{\varTheta}^{k}$ : Apply the operation \eqref{Dq}
		
		\STATE $\overline{B}^k\gets\overline{\mathcal{B}}^k$ : IDCT($\overline{\mathcal{B}}^k$) according to \eqref{IDCT}
		
		\ENDFOR
		
		\STATE $\mathcal{S}\gets \overline{B}^{1}\cup\overline{B}^{2}\cup\cdots\cup\overline{B}^{k}\cup\cdots $
		
		\RETURN $\mathcal{S}$
	\end{algorithmic} 
\end{algorithm}

\subsection{Extracting algorithm}

The detailed process of extracting the secret message from the stego image $\mathcal{S}$ is described below. The step-by-step procedure is included in Algorithm \ref{alg:extracting}. Taking as input the stego image $\mathcal{S}$, this is divided into non-overlapping $8\times8$ bytes blocks. Afterwards, DCT coefficients are obtained from each $8\times8$ blocks, see \eqref{DDCT}. The quantized DCT coefficients from each block using quantization matrix \eqref{QTable} are obtained. Then, the first eight AC coefficients, are collected from all $8\times8$ DCT coefficient matrices and stored into a coefficient 1D-array $\omega$. The zigzag scan order is used to collect the coefficients, see Figure \ref{ZZagF}. In order to increase the security of the stego system, a 128-bit private key $\kappa$ is used. From this a 1024-bit sequence is generated using BLAKE2B. Then, this sequence is expanded to a binary sequence $\overline{\kappa}$ with the same length of $\omega $. Next, $\omega =\omega^{(1)}||\omega^{(2)}||\cdots||\omega^{(i)}||\cdots $ and $\overline{\kappa}=\overline{\kappa}^{(1)}||\overline{\kappa}^{(2)}||\cdots||\overline{\kappa}^{(i)}||\cdots$ are partitioned into logical sequences of length 64. Then, for each $\omega^{(i)}$ the LSB of the value of the coefficient of a particular position $\rho_i$ is extracted and appended to form the final bits stream $\M$ of the secret message. The $\rho_i$ positions are determined by $\overline{\kappa}^{(i)}$ and chaotic positions generator $(\varrho_i)_{1\leq i\leq 64}$ given by \eqref{fcmg}, see Algorithm \ref{Permutation}.

\begin{algorithm}[H]
	\caption{Extracting Algorithm}
	\label{alg:extracting}
	\begin{algorithmic}[1]	
		\STATE \text{{\bf Input:}} Stego image $\Steg$, 128-bit private key $\kappa$,  $x(0)$, $\nu$	
		\STATE \text{{\bf Output:}} Secret message $\M$,\\
		\mbox{\textbf{/* Collecting AC coefficients */}}
		\STATE Divide $\Steg$ into $\mathcal{K}$ non-overlapping blocks of $8\times8$ bytes, see \eqref{Cover_Bloks}
		
		\STATE $\omega=\emptyset$
		
		\STATE $\ell=0$
		
		\FOR{\textbf{each} $B^k\in \Steg$}
		
		\STATE $\mathcal{B}^k\gets B^k$ : DCT($B^k$) according to~\eqref{DDCT}
		J. Inst. Eng. India Ser. B
		\STATE $\varTheta^{k} \gets \mathcal{B}^{k}$		
		: Quantify $\mathcal{B}^{k}$ according to~\eqref{quantification}
		
		\STATE $\nu^{k} \gets \varTheta^{k}$ : Apply the zigzag scan, see Figure \ref{ZZagF}
		
		\STATE $\omega\gets\omega\cup\{\nu_j^{k}\}_{2\leq j\leq 9} $
		
		\ENDFOR\\
		\mbox{\textbf{/* Key expansion using blake2b algorithm */}}
		
		\STATE From $\kappa$ generate 1024-bit sequence using BLAKE2B algorithm
		
		\STATE Expand the previous sequence to a binary sequence $\overline{\kappa}$ with the same length of $\omega$\\
		\mbox{\textbf{/* Extracting secrete bits */}}
		
		\STATE Partition $\omega \gets\omega^{(1)}||\omega^{(2)}||\cdots||\omega^{(i)}||\cdots $ and $\overline{\kappa}\gets\overline{\kappa}^{(1)}||\overline{\kappa}^{(2)}||\cdots||\overline{\kappa}^{(i)}||\cdots$ into a logical sequence of length 64
		
		\FOR{\textbf{each} $\omega^{(i)}\in\omega$}
		
		\STATE $\rho\gets\mbox{ChaoticPositions}(\overline{\kappa}^{(i)}, x(0), \nu)$
		
		\FOR{\textbf{each} $\psi\in\rho$}
		
		\STATE $\ell\gets\ell + 1$
		
		\STATE $m_{\ell}\gets $R$^{-1}(\mbox{abs}(\omega^{(i)}_{\psi}))$
		
		\ENDFOR
		
		\ENDFOR		
		
		\RETURN $\mathcal{\M}$
	\end{algorithmic} 
\end{algorithm}

\section{Results and discussion}

\begin{figure}[H]
	\centering\includegraphics[scale=.49]{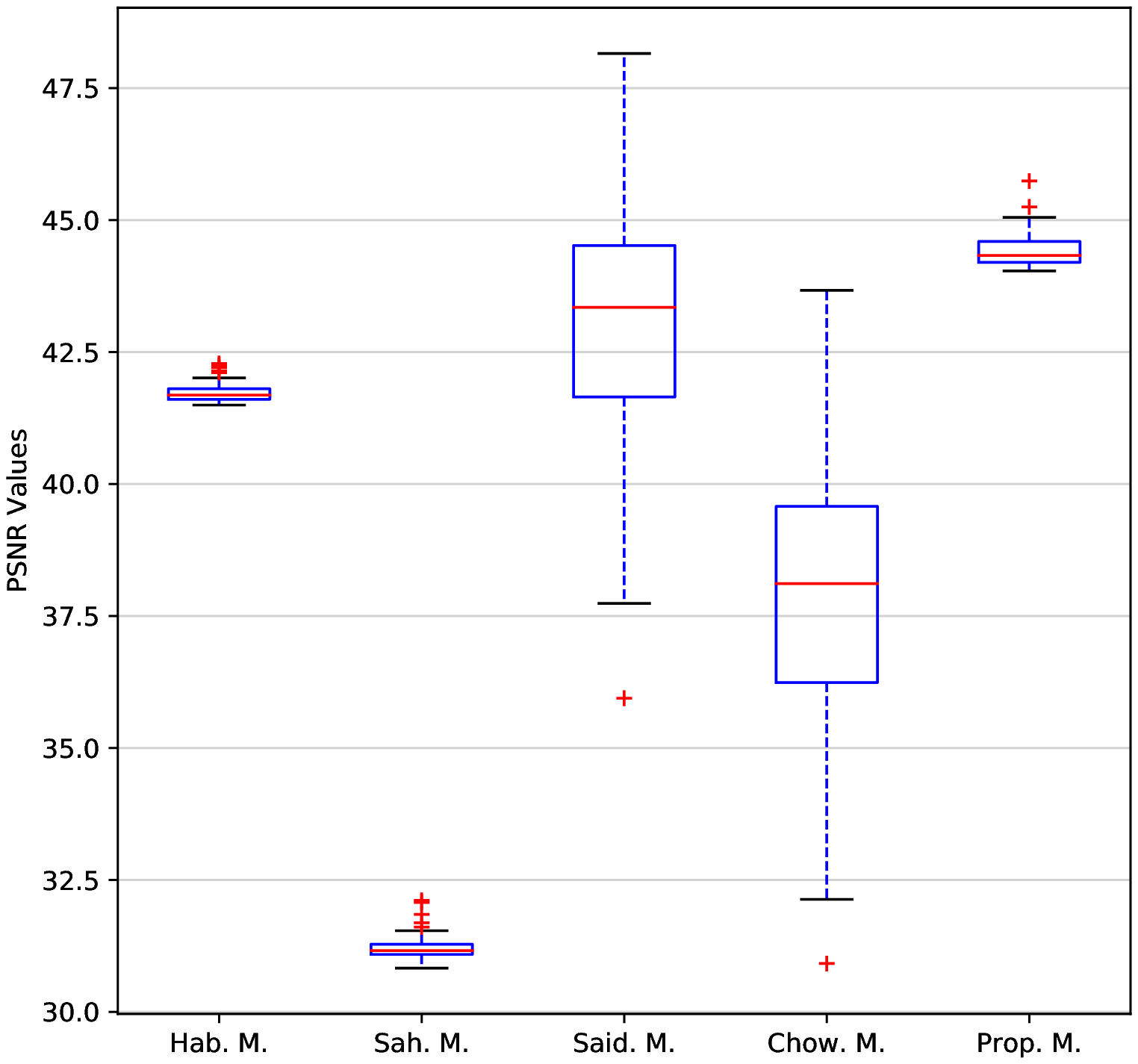}
	\centering\includegraphics[scale=.49]{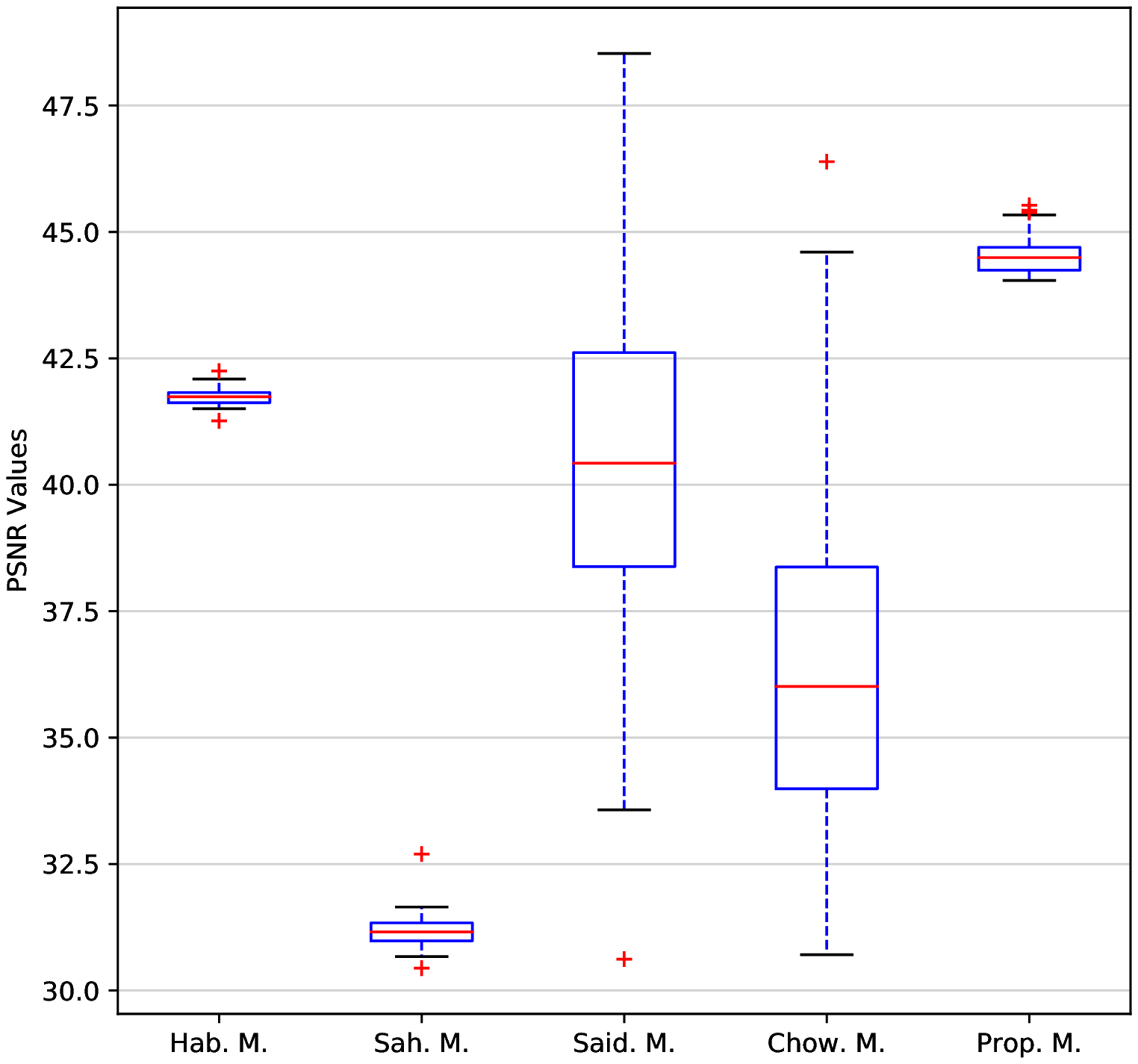}
	\caption[]{PSNR values. The first row contains the PSNR values corresponding to the first dataset while the another to second}
	\label{PSNR_DS-I-IV}	
\end{figure}

The results showed that proposed method reached higher PSNR values (75$\%$) than the  Habib et al., Sahar et al., Saidi et al. and Chowdhuri et al. methods (Figure \ref{PSNR_DS-I-IV}), exhibiting higher PNSR values of 42.5 db.  This indicates that obtained stego images by proposed method have higher imperceptibility level compared to the other methods. This is consistent with those results of PSNR (30 to 50 db) found by \cite{Coskun}.

\begin{figure}[H]
	\centering\includegraphics[scale=.49]{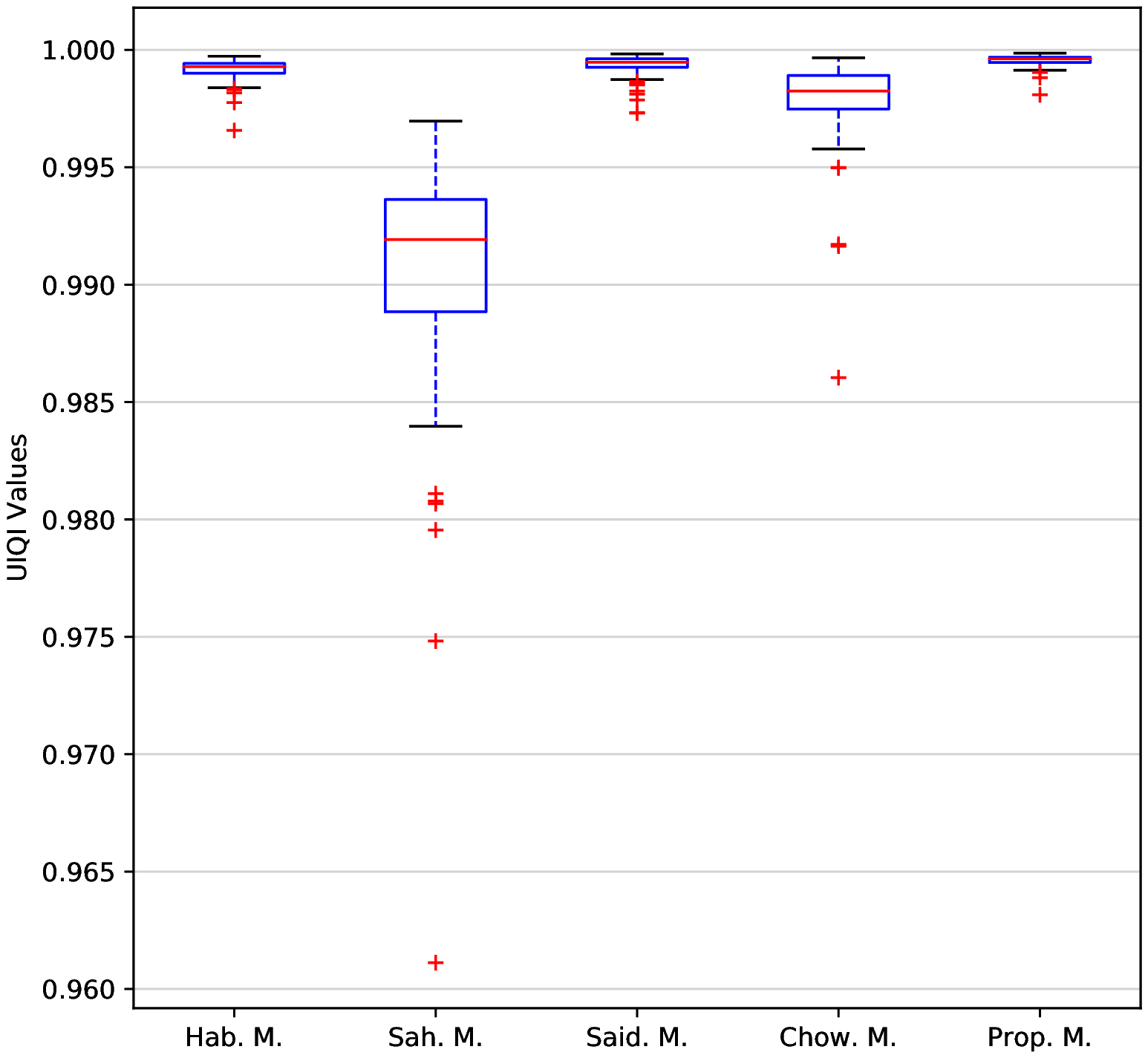}
	\centering\includegraphics[scale=.49]{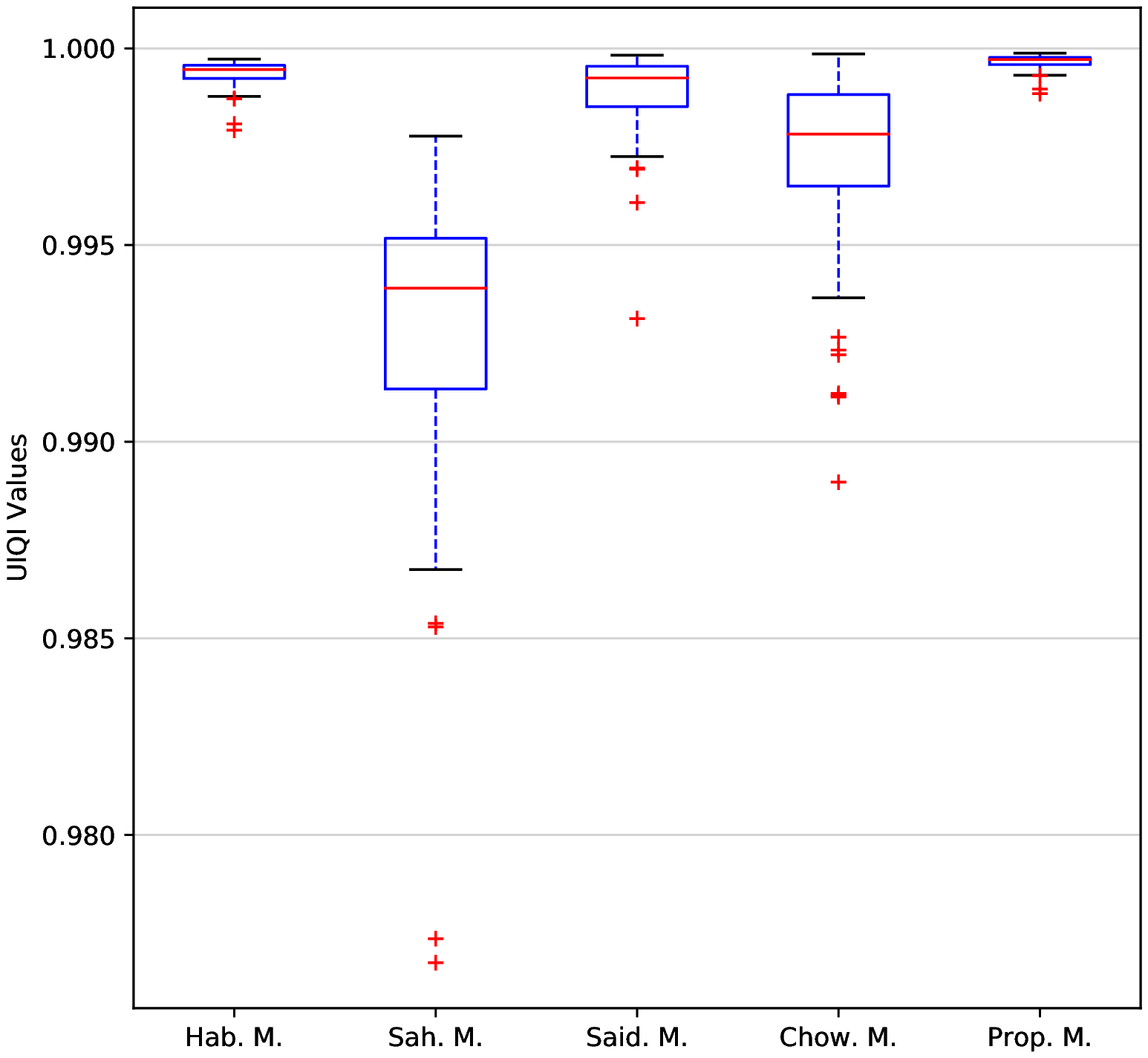}
	\caption[]{UIQI values} 
	\label{UIQI-V}	
\end{figure}

Image quality was significantly improved (75$\%$) by the proposed method in comparison with the other methods (Figure \ref{UIQI-V}). This indicates that the cover image quality is not significantly different from the stego images due to UIQI values are close to the unit. Also, it resulted in obtained stego images had a good visual quality with respect to the other methods.

\begin{figure}[H]
	\centering\includegraphics[scale=.49]{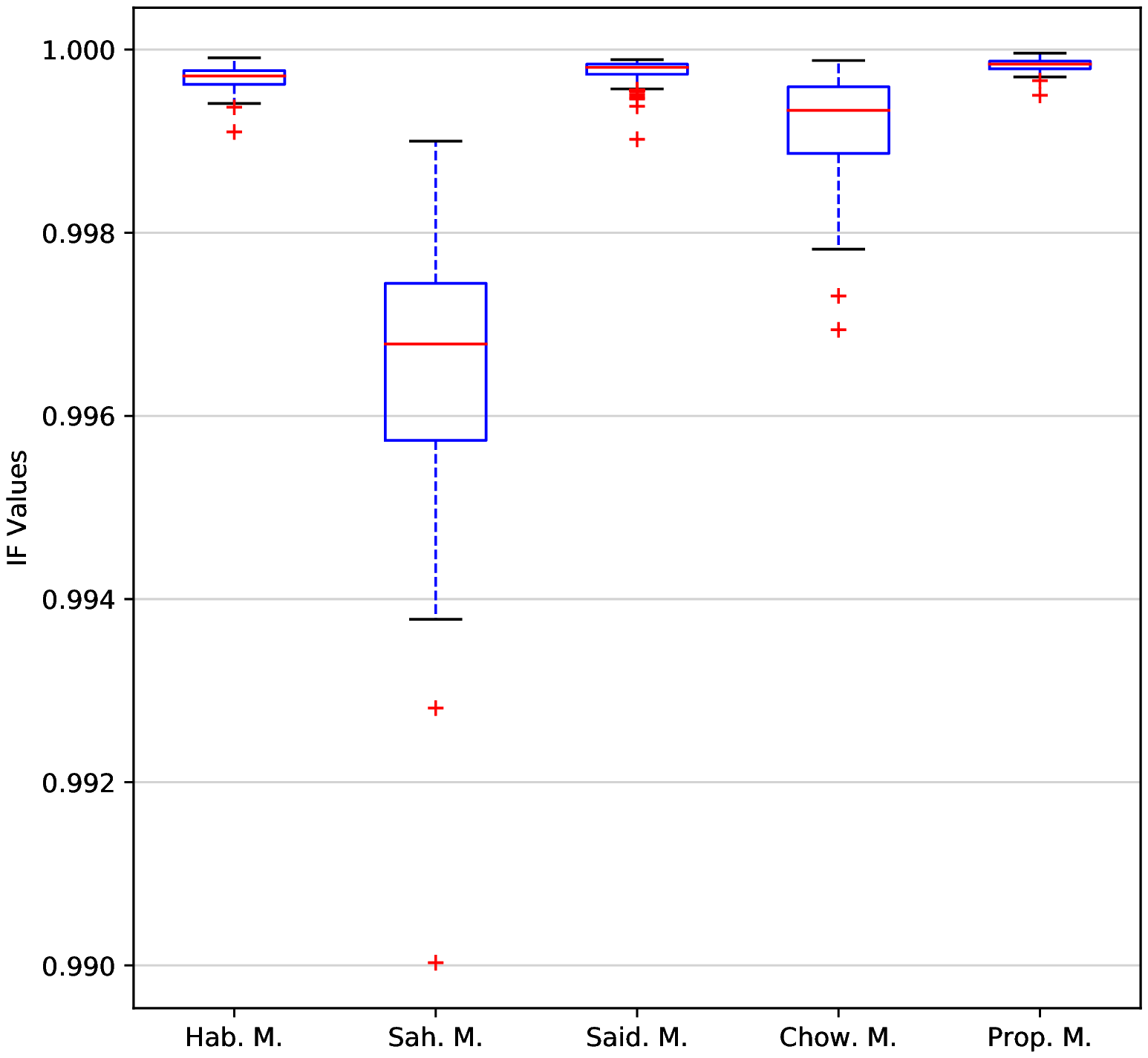}
	\centering\includegraphics[scale=.49]{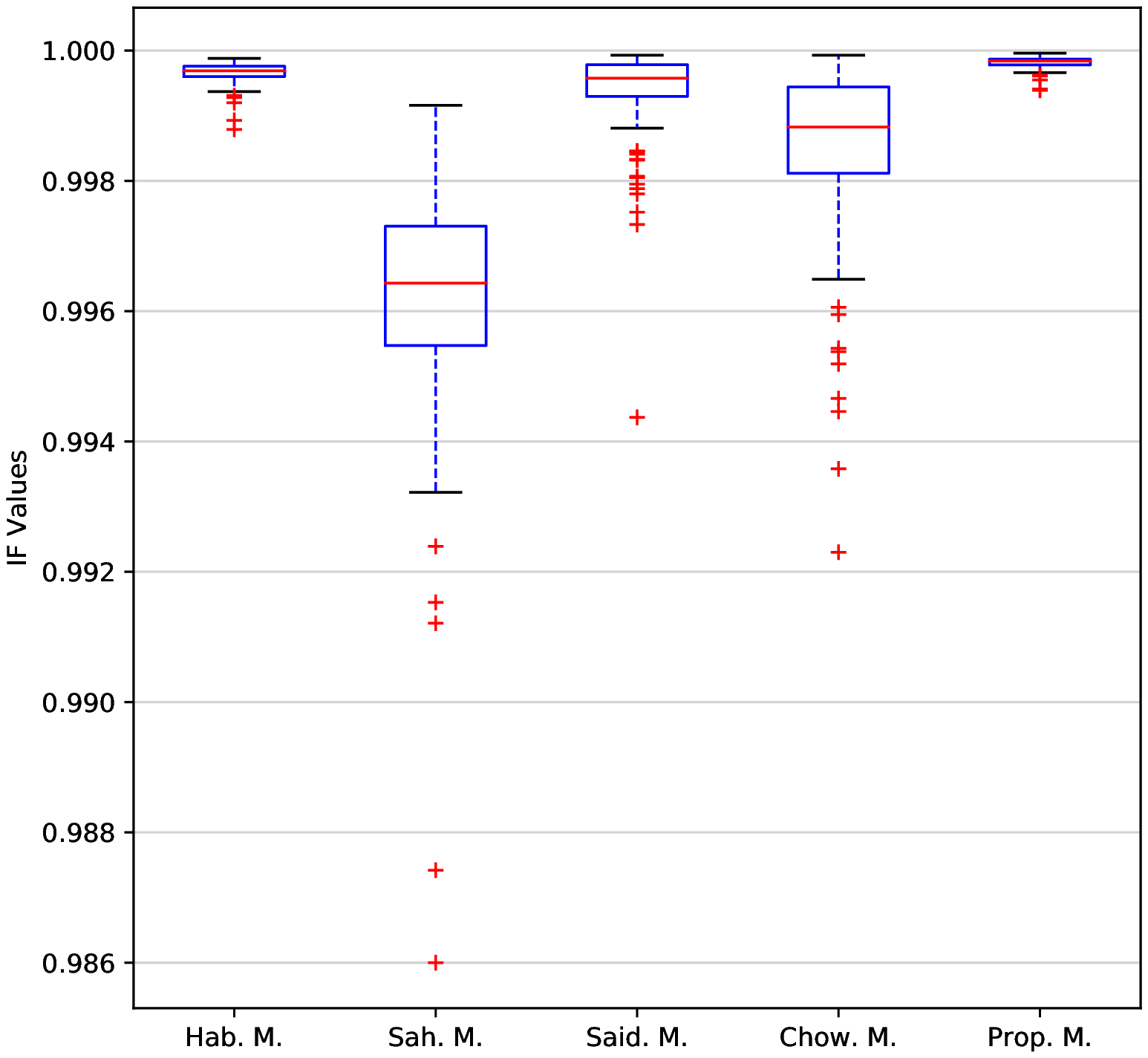}
	\caption[]{IF values} 
	\label{IF-V}	
\end{figure}

The proposed method enhanced notably the image fidelity (75$\%$) compared to the other methods (Figure \ref{IF-V}). This means that the cover images have the similarity high level compared to the stego images. This results are consistent with those obtained by \cite{SoriaSCN} who found IFh values close to the unit.

\begin{figure}[H]
	\centering\includegraphics[scale=.49]{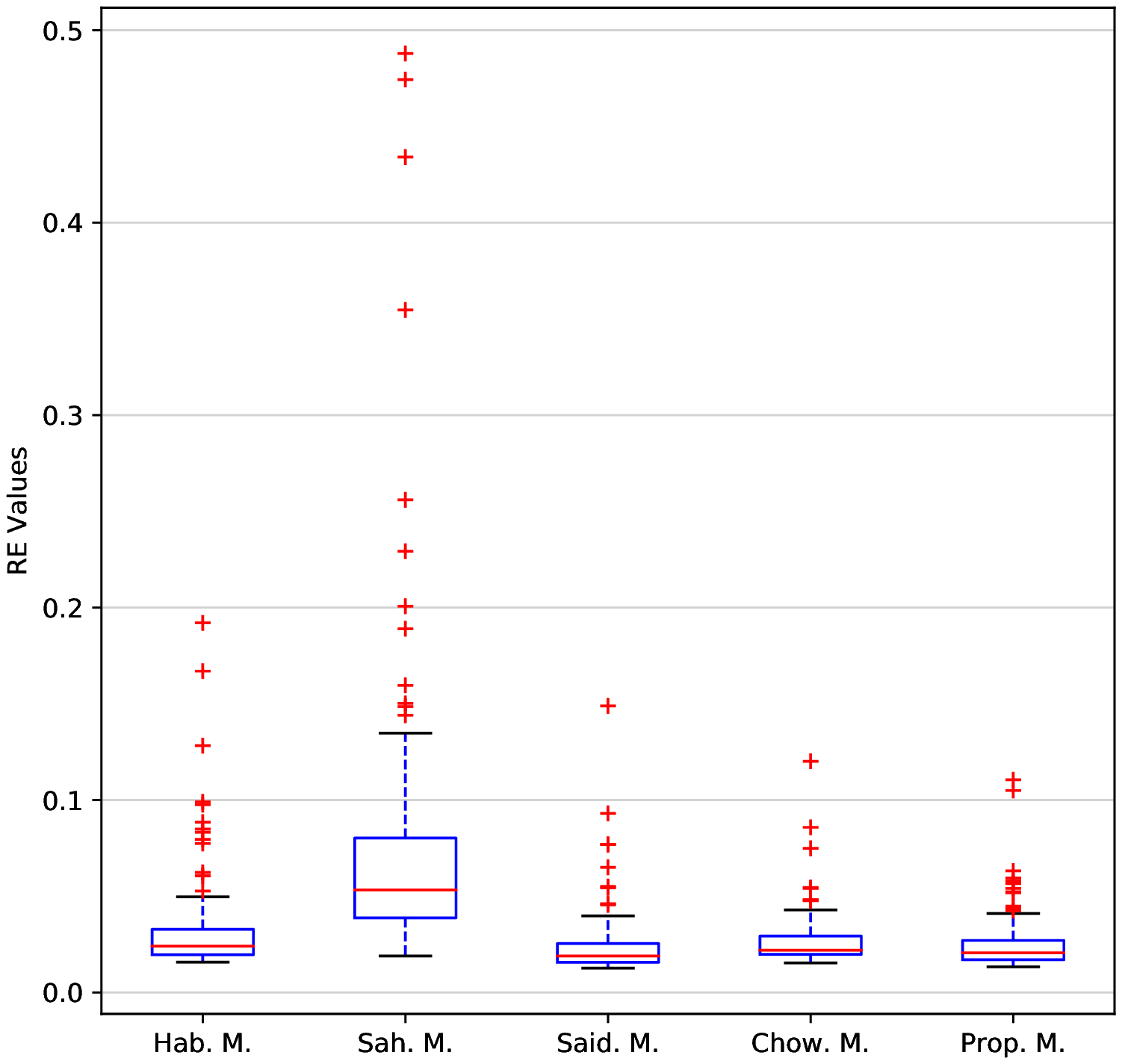}
	\centering\includegraphics[scale=.49]{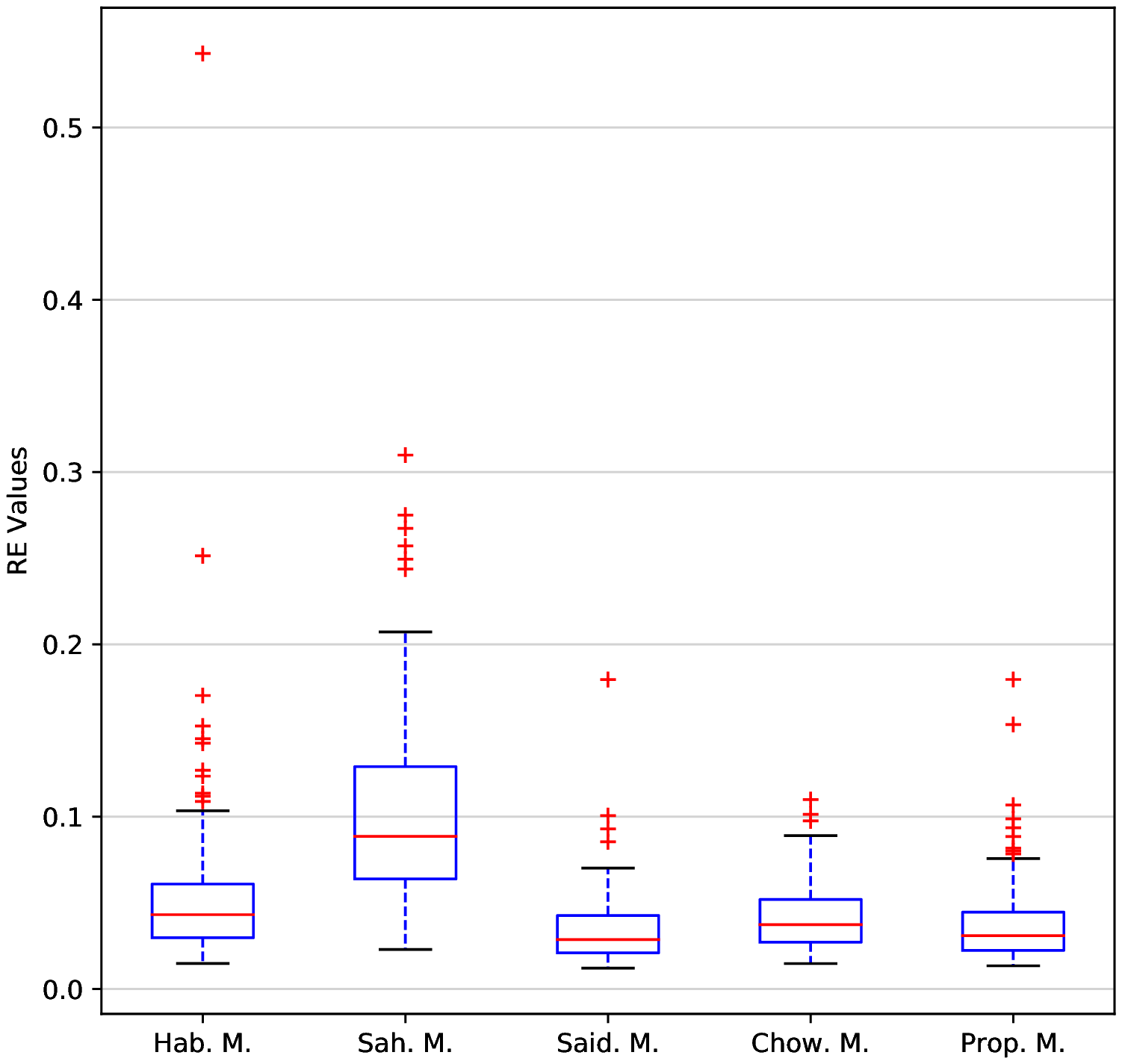}
	\caption[]{RE values} 
	\label{RE_I-IV}	
\end{figure}

The ER values were slightly better in the proposed method than those obtained by the methods of Habib et al., Sahar and Chowdhuri et al. However, Saidi et al. method showed smaller RE values than proposed method (Figure \ref{RE_I-IV}). Decreased RE values resulted in steganographic system that is sufficiently safe to establish a private and confidential communication between two parts \cite{Cachin,SoriaSCN}.

\section{Conclusions}

We have presented a steganographic algorithm embedding a secrete message in the first eight AC coefficients. Experimental analysis of two datasets for stoganographic images revealed that 100$\%$ of the PSNR values of the proposed method are greater than the  PSNR values of the state-of-art analyzed, except for Saidi et al. method. The method proposed improved notably image quality, imperceptibility and similarity and provided high imperceptibility level and good embedding capacity compared to existing state-of-the-art methods. We expect the presented method can protect the information and establish a safe communication between parts.

%


\end{document}